\documentclass[prb,aps,twocolumn,groupedaddress]{revtex4}

\usepackage{bm}
\usepackage{graphicx}
\usepackage{color}
\usepackage{ulem}

\begin{document}

\title{Dichotomy of superconductivity between monolayer FeS and FeSe}

\author{Koshin Shigekawa$^1$, Kosuke Nakayama$^{1,2}$, Masato Kuno$^1$, Giao N. Phan$^1$, Kenta Owada$^1$, Katsuaki Sugawara$^{1,3,4}$, Takashi Takahashi$^{1,3,4}$, and Takafumi Sato$^{1,3,4}$}

\affiliation{$^1$Department of Physics, Tohoku University, Sendai 980-8578, Japan\\
$^2$Precursory Research for Embryonic Science and Technology, Japan Science and Technology Agency, Kawaguchi, Saitama 332-0012, Japan\\
$^3$Center for Spintronics Research Network, Tohoku University, Sendai 980-8577, Japan\\
$^4$WPI Research Center, Advanced Institute for Materials Research, Tohoku University, Sendai 980-8577, Japan\\
}

\begin{abstract}
The discovery of high-temperature ($T_{\rm c}$) superconductivity in monolayer FeSe on SrTiO$_3$ raised a fundamental question whether high $T_{\rm c}$ is commonly realized in monolayer iron-based superconductors. Tetragonal FeS is a key material to resolve this issue because bulk FeS is a superconductor with $T_{\rm c}$ comparable to that of isostructural FeSe. However, difficulty in synthesizing tetragonal monolayer FeS due to its metastable nature has hindered further investigations. Here we report elucidation of band structure of monolayer FeS on SrTiO$_3$, enabled by a unique combination of in-situ topotactic reaction and molecular-beam epitaxy. Our angle-resolved photoemission spectroscopy on FeS and FeSe revealed marked similarities in the electronic structure, such as heavy electron doping and interfacial electron-phonon coupling, both of which have been regarded as possible sources of high $T_{\rm c}$ in FeSe. However, surprisingly, high-$T_{\rm c}$ superconductivity is absent in monolayer FeS. This is linked to the weak superconducting pairing in electron-doped multilayer FeS in which the interfacial effects are absent. Our results strongly suggest that the cross-interface electron-phonon coupling enhances $T_{\rm c}$ only when it cooperates with the pairing interaction inherent to the superconducting layer. This finding provides a key insight to explore new heterointerface high-$T_{\rm c}$ superconductors.\end{abstract}
\maketitle

Tetragonal FeSe is an iron-based superconductor with the simplest layered structure, and has received tremendous attention because high-temperature ($T_{\rm c}$) superconductivity was discovered in one-monolayer (1-ML) film grown on SrTiO$_3$ substrate (1). The $T_{\rm c}$ value of 1-ML FeSe ($\sim$ 65 K) (2,3) is almost an order of magnitude higher than that ($\sim$8 K) of bulk FeSe (4), and is the highest among iron-based superconductors. The mechanism of such $T_{\rm c}$ enhancement is a central issue in condensed matter physics. The widely discussed scenario is that the high $T_{\rm c}$ is closely linked to the coupling of FeSe layer with SrTiO$_3$ substrate (5). A major role of SrTiO$_3$ substrate is an electron doping to the FeSe layer, producing a large electronlike Fermi surface centered at the M point in the Brillouin zone (2,3), in sharp contrast to bulk FeSe where hole and electron pockets coexist at the $\Gamma$ and M point, respectively (6,7). While this excess electron doping seems to enhance $T_{\rm c}$, the maximum $T_{\rm c}$ achieved solely by electron doping to multilayer FeSe thin films (8,9) never reaches 65 K. Therefore, substrate must play another important role to enhance $T_{\rm c}$ in 1-ML FeSe. As such, a strong coupling of SrTiO$_3$ phonons with FeSe electrons across the interface has been proposed (10-12). This effect has been observed as the replica band separated by the optical phonon energy from the main band (10-12). However, it remains still elusive whether the electron-phonon coupling alone is sufficient to account for the high $T_{\rm c}$ (13-18). A crucial test to clarify the contribution of interfacial electron-phonon coupling is a comparative study on a 1-ML film of other iron-based superconductors and to elucidate whether a similar high-$T_{\rm c}$ superconductivity associated with the electron-phonon coupling can be realized. Thus, the synthesis of such 1-ML system is of great importance for a crucial understanding on the origin of high-$T_{\rm c}$ superconductivity. Also, this study would provide a pathway to further exploration of new high-$T_{\rm c}$ systems.

\begin{figure*}
\includegraphics[width=5.5in]{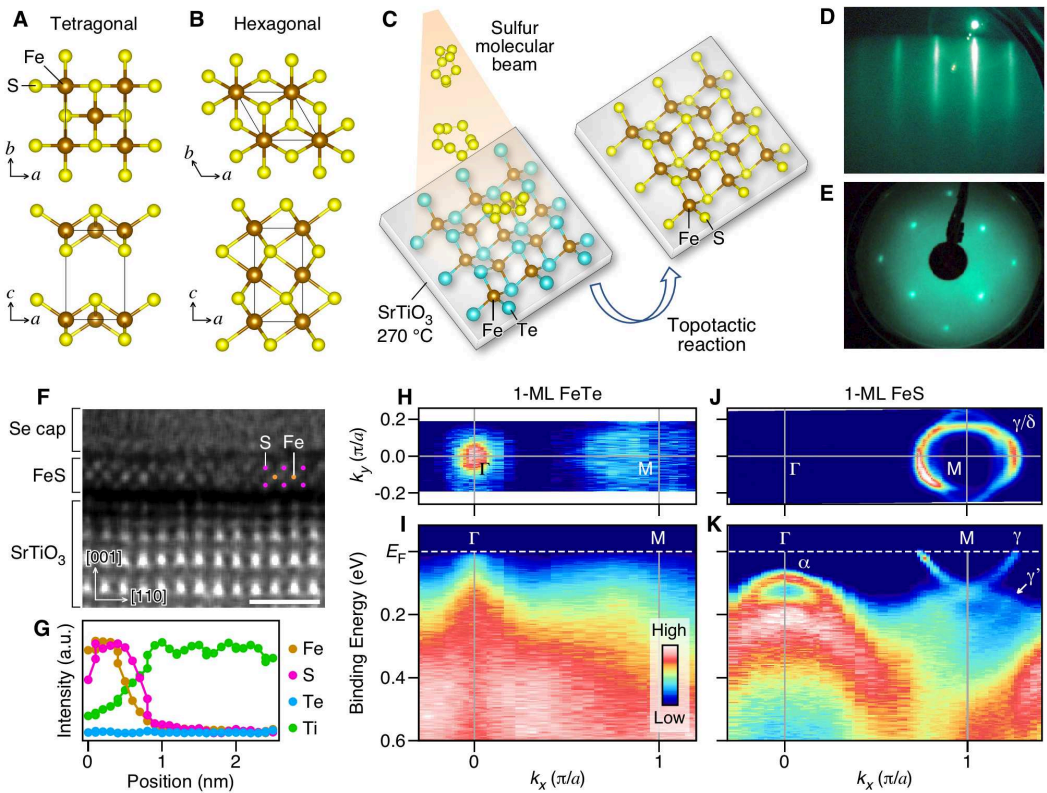}
\vspace{0cm}
\caption{Topotactic synthesis of 1-ML tetragonal FeS film. (A and B) Crystal structures of tetragonal and hexagonal FeS, respectively. (C) Schematic for topotactic growth process of 1-ML tetragonal FeS film on SrTiO$_3$ substrate. (D) Reflection high-energy electron diffraction (RHEED) pattern of 1-ML FeS. (E) Low-energy electron diffraction (LEED) pattern of 1-ML FeS measured with an incident electron energy of 70 eV. (F) High-angle annular dark-field scanning transmission electron microscopy (HAADF-STEM) image viewed along the 110 direction of Se-capped FeS film on SrTiO$_3$. White scale bar, 1 nm wide. (G) Composition line profile across the interface measured by energy-dispersive x-ray spectroscopy (EDX) plotted with normalized intensity. (H) ARPES intensity mapping at $T$ = 30 K for 1-ML FeTe film plotted as a function of two-dimensional wave vector measured with the He-I$\alpha$ line. The intensity is obtained by integrating the spectral intensity within $\pm$10 meV of $E_{\rm F}$. (I) ARPES intensity plot as a function of binding energy and wave vector measured along the $\Gamma$M line. (J and K) Same as (H and I), respectively, but for 1-ML FeS film.}
\end{figure*}

Tetragonal FeS and FeTe are particularly suited to resolve this issue because they are isostructural to FeSe. However, in FeTe, the robust antiferromagnetism (19) seems to kill the superconductivity, so that the high-$T_{\rm c}$ superconductivity has not been observed in either bulk or 1-ML film on SrTiO$_3$ (a signature of interfacial electron-phonon coupling is also absent) (20). On the other hand, it was recently reported that FeS shows a bulk superconductivity with $T_{\rm c}$ = 4.5 K (21), comparable to that of bulk FeSe, providing a hopeful platform to explore the high-$T_{\rm c}$ superconductivity in the 1-ML system. However, a high-quality tetragonal FeS thin film (Fig. 1A) is known to be hardly obtained by the molecular-beam epitaxy (MBE) technique which is widely used for synthesizing 1-ML FeSe and FeTe, and hence the presence or absence of high-$T_{\rm c}$ superconductivity as well as the interfacial effects has not been verified (22,23). This is mainly due to the high vapor pressure of S atoms, and more critically, to the existence of the hexagonal phase (Fig. 1B) that is thermodynamically more stable than the tetragonal phase. Despite these difficulties, it is strongly desired to find a new route to synthesize a high-quality tetragonal FeS film.

Here we demonstrate an efficient synthesis route of high-quality tetragonal 1-ML FeS. We adopted the topotactic reaction, which can convert the crystalline phase with keeping the lattice symmetry of parent crystals. Using this method with FeTe as a starting material, we have successfully grown a 1-ML tetragonal FeS film on a SrTiO$_3$ substrate. By angle-resolved photoemission spectroscopy (ARPES), we have revealed that 1-ML FeS definitely experiences an interfacial effect. But, to our surprise, 1-ML FeS does not show high-$T_{\rm c}$ superconductivity in sharp contrast to FeSe. We show that the dichotomy of superconductivity between FeS and FeSe monolayers is the key to understand the high $T_{\rm c}$ mechanism of Fe-chalcogenides.

\begin{figure*}
\includegraphics[width=5.5in]{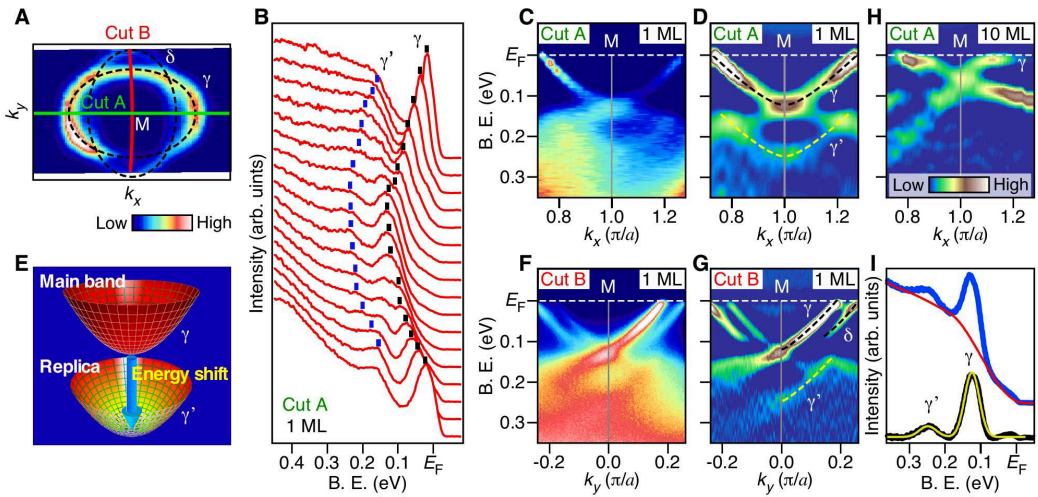}
\vspace{0cm}
\caption{Interfacial electron-phonon coupling in 1-ML FeS. (A) A magnified view of the Fermi surface around the M point of 1-ML FeS film. (B and C) Near-$E_{\rm F}$ APRES spectra and its intensity plot along cut A in (A), respectively, measured at $T$ = 30 K. Black and blue dots in (B) are a guide for the eyes to trace the main and replica bands, respectively. (D) Second derivative intensity plot of (C). Black and yellow dashed curves are a guide for the eyes to trace the main and replica bands, respectively. (E) Schematics for the main and replica bands. (F and G) Same as (C and D), respectively, but measured along cut B in (A). (H) Same as (D), but for 10-ML FeS film. (I) Raw ARPES spectrum at the M point extracted from (F) (blue curve), the background used in fitting (red), background-subtracted ARPES spectrum (black), and the result of fitting with two Lorentzian curves (yellow). The background was modelled by a spline interpolation by following previous studies (9-11).
}
\end{figure*}

As schematically shown in Fig. 1C, the topotactic substitution of Te atoms with S atoms is promoted by heating a MBE-grown FeTe film at 270$^{\circ}$C while showering S molecular beams onto the film surface in a vacuum for 5 min (see Methods for details of the FeTe-growth and sulfurization conditions). The resultant film has a clean and flat surface, as recognized from a sharp streak RHEED (reflection high-energy electron diffraction) pattern in Fig. 1D. Importantly, the tetragonal lattice is maintained, as evidenced by the four-fold symmetry of the LEED (low-energy electron diffraction) pattern (Fig. 1E). In addition, STEM (scanning transmission electron microscopy) measurements revealed three atomic layers (Fig. 1F), consistent with the formation of a 1-ML film. Furthermore, we performed EDX (energy-dispersive x-ray spectroscopy) measurements to estimate the S content in the film. As visible in Fig. 1G, the composition line profile signifies the presence of S atoms with no trace of Te atoms, showing an almost 100$\%$ substitution of Te atoms with S atoms. This is corroborated by the absence of Te-derived core-level peaks in the XPS (x-ray photoemission spectroscopy) spectrum (SI Appendix Fig. S1). All these characterizations unambiguously demonstrate a successful synthesis of tetragonal 1-ML FeS film on SrTiO$_3$ by in-situ topotactic reaction. It is noted that the spatial resolution of the EDX analysis in the present experimental setup is estimated to be about 2 nm. This is comparable to the full probed width of FeS/SrTiO$_3$ ($\sim$2.5 nm, Fig. 1G), meaning that even a sharp step edge in the composition is substantially broadened in the EDX line profile due to the spatial resolution. Thus, we think that the apparent diffusion of Ti in the EDX line profile is due to the finite spatial resolution in the EDX measurement.

The conversion from FeTe to FeS is also confirmed by a comparison of the electronic structure by ARPES (Fig. 1 H-K). As seen from Fig. 1 H and I, an as-grown 1-ML FeTe film has a holelike Fermi surface centered at the $\Gamma$ point. In addition, a significant spectral broadening is observed as in the previous study (24), possibly due to the strong correlation effect (25). In contrast, sulfurization leads to a development of sharp quasiparticle peaks (Fig. 1K), enabling a precise determination of the electronic structure: the top of the holelike band ($\alpha$) at the $\Gamma$ point is located at $\sim$80 meV below the Fermi level ($E_{\rm F}$), and the Fermi surface consists only of large electron pockets ($\gamma$/$\delta$) centered at the M point (Fig. 1 J and K). The observed drastic change in the electronic structure strongly supports our successful sulfurization (we found that the same method works also for the conversion from FeTe to FeSe and from FeSe to FeS; SI Appendix Fig. S2). Another important finding is that the electron carrier concentration estimated from the Fermi-surface volume in Fig. 1J is very high ($\sim$0.15 electrons per Fe), indicating that the heavy electron doping takes place in 1-ML FeS. This would be a result of charge transfer from the SrTiO$_3$ substrate, similarly to the case of 1-ML FeSe (2,3). We verified the charge transfer by performing thickness-dependent ARPES measurements (SI Appendix Fig. S3). It is worth noting that the doping level of 0.15 electrons/Fe is well inside the high-$T_{\rm c}$ phase in the case of 1-ML FeSe (26).

A careful look at the ARPES intensity further reveals an intriguing subband feature ($\gamma$) indicated by a white arrow in Fig. 1K. To discuss the origin of this band in more detail, we show in Fig. 2 the ARPES spectra and intensity measured around the M point. Along cut A in Fig. 2A, one can clearly see the $\gamma$ and $\gamma$' bands (Fig. 2 B-D). The observed $\gamma$'-band dispersion is well reproduced by an approximately 100-meV downward shift of the $\gamma$-band dispersion (see dots in Fig. 2B, dashed curves in Fig. 2D, and schematic in Fig. 2E), and the energy separation between the $\gamma$' and $\gamma$ bands is close to the energy of optical phonons in SrTiO$_3$. These characteristics of the $\gamma$' band are qualitatively similar to those of the main-band replica produced by the phonon shake-off effect in 1-ML FeSe (10-12). Therefore, the $\gamma$' band is likely attributed to the replica band caused by the coupling with SrTiO$_3$ phonons. It is noted that a possibility that the replica band is simply the outer electron pocket (27) is ruled out because the $\gamma$' band is not connected to the outer electron pocket ($\delta$) as observed in cut B (Fig. 2 F and G). The interface origin of the $\gamma$' band is also supported by the fact that it disappears in multilayer films (Fig. 2H and SI Appendix Fig. S4).

\begin{figure*}
\includegraphics[width=5.5in]{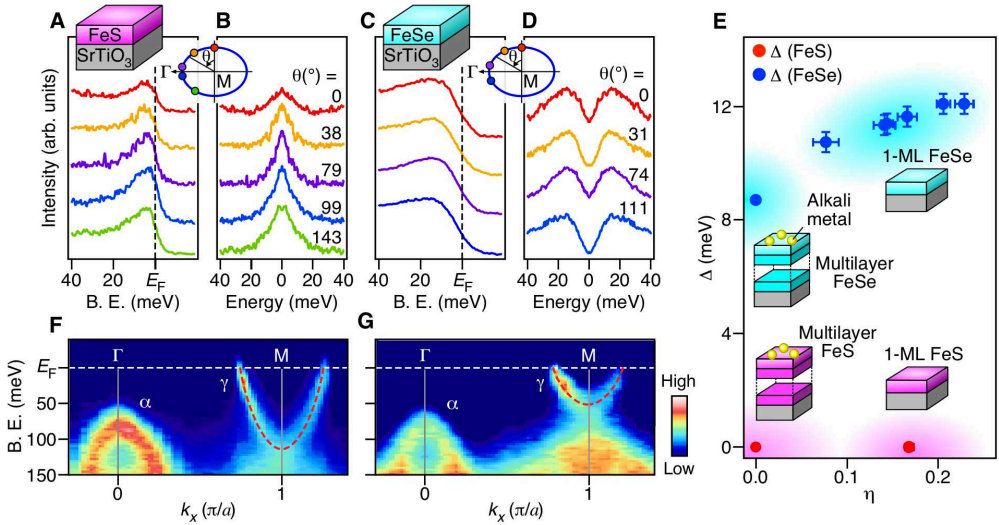}
\vspace{0cm}
\caption{Absence of high-$T_{\rm c}$ superconductivity in 1-ML FeS. (A and B) Near-$E_{\rm F}$ ARPES spectra and corresponding symmetrized spectra (obtained by symmetrizing the spectrum with respect to $E_{\rm F}$ to eliminate the effect of Fermi-Dirac distribution function), respectively, measured at $T$ = 10 K at the $k_{\rm F}$ points of electron pocket. The inset shows a schematic view of 1-ML FeS on SrTiO$_3$ and an electron Fermi surface together with the definition of Fermi-surface angle ($\theta$). Filled circles on the Fermi surface indicate the momentum location where the ARPES spectra in (A and B) were obtained, and their colorings correspond to those of the spectra. (C and D) Same as (A and B), respectively, but for 1-ML FeSe on SrTiO$_3$. (E) Experimentally-determined $\Delta$ for 1-ML FeS, 1-ML FeSe (11), alkali-metal-dosed multilayer FeS, and alkali-metal-dosed multilayer FeSe (7), plotted as a function of $\eta$ (the ratio of the $\gamma$'-band intensity to the $\gamma$-band intensity). (F and G) Band dispersions along the $\Gamma$M line in 1-ML FeS and FeSe, respectively. Red dashed curves are a guide for the eyes to trace the electron pocket.}
\end{figure*}

To estimate the strength of electron-phonon coupling, we performed numerical fittings to the ARPES spectrum at the M point (Fig. 2I) by following the previous studies on FeSe that proposed the intensity ratio of $\gamma$' to $\gamma$ band ($\eta$= I$_{\gamma'}$/I$_{\gamma}$) to be proportional to the coupling constant (10-12). The obtained $\eta$ of 0.17 is comparable to that reported for 1-ML FeSe (10-12), suggesting that the interfacial electron-phonon coupling in 1-ML FeS is as strong as that in 1-ML FeSe. From the numerical fittings, we also estimated the precise energy separation between the $\gamma$ and $\gamma$' bands to be 110 meV, which is comparable to that in 1-ML FeSe ($\sim$100 meV).

Having established the presence of the electron doping and the interfacial electron-phonon coupling, it is tempting to search for high-$T_{\rm c}$ superconductivity in 1-ML FeS. Fig. 3A shows high-resolution ARPES spectra measured at several Fermi wave vector ($k_{\rm F}$) points of the $\gamma$ band at $T$ = 10 K. One can immediately see that the leading-edge midpoint is always located at around $E_{\rm F}$ irrespective of the $k_{\rm F}$ points. Corresponding symmetrized spectra in Fig. 3B show a single peak at $E_{\rm F}$, indicative of the absence of a superconducting gap. This suggests that 1-ML FeS does not show high-$T_{\rm c}$ superconductivity, in sharp contrast to 1-ML FeSe (Fig. 3 C and D) where the ARPES spectrum obviously shows a depletion of spectral weight at $E_{\rm F}$. We also found that the multilayer FeS films neither show high-$T_{\rm c}$ superconductivity even when electrons are sufficiently doped (SI Appendix Fig. S4), again in contrast to the case of K-coated multilayer FeSe with the $T_{\rm c}$ over 40 K (8,9).

The observed contrasting behavior between the electron-doped FeS and FeSe thin films are summarized in Fig. 3E, where we see that even in the absence of interfacial electron-phonon coupling ($\eta$ = 0, corresponding to the alkali-metal-coated multilayer films), FeSe shows high-$T_{\rm c}$ superconductivity (40-48 K) with a superconducting-gap size $\Delta$ of $\sim$8 meV (8,9), while in FeS the magnitude of $T_{\rm c}$ and $\Delta$ appears to be below our detectable limits (SI Appendix Fig. S4). It has been discussed that an inclusion of the interfacial electron-phonon coupling ($\eta$ $>$ 0, corresponding to 1-ML films) leads to an enhancement in $T_{\rm c}$ and $\Delta$ by about 20 K and 5 meV, respectively in FeSe (2,3), while in FeS such an enhancement in $T_{\rm c}$ or $\Delta$ is not observed (Fig. 3 A and B).

One may think that the absence of high-$T_{\rm c}$ superconductivity in 1-ML FeS may be due to some extrinsic effects such as disorder in the crystal. We have examined the possibility of disorder in the FeS film from the ARPES measurements. It is well known that the disorder in the crystal causes a considerable broadening of quasiparticle peaks in the ARPES spectrum due to the shortened quasiparticle lifetime. If a substantial disorder is present in our 1-ML FeS and suppresses the superconductivity, the quasiparticle peak would be broader than that of superconducting 1-ML FeSe. However, as seen in Fig. 3A and C, the quasiparticle peak in 1-ML FeS is much sharper than high-$T_{\rm c}$ 1-ML FeSe. This indicates that our 1M-FeS has a negligible disorder in the crystal. Since one may also wonder whether the lattice strain may suppress the superconductivity, we compared the in-plane lattice constant extracted from the LEED pattern. The results give $a$ = 3.87$\pm$0.04 \AA \space for 1-ML FeS on SrTiO$_3$, which is about 5$\%$ larger than that of bulk FeS (3.68 \AA), suggesting that there is a tensile strain in the 1-ML FeS film due to the lattice mismatch with the SrTiO$_3$ substrate ($a$ = 3.91 \AA). Since high-$T_{\rm c}$ superconductivity occurs in 1-ML FeSe with a similar in-plane lattice constant ($a$ $\sim$ 3.9 \AA) or a similar magnitude of tensile strain ($\sim$5$\%$) (3,28), the absence of high-$T_{\rm c}$ superconductivity in 1-ML FeS would not be directly linked to the tensile strain.

The observed striking difference between FeS and FeSe rather suggests that, while the electron doping and the resultant change in the Fermi-surface topology may be important to realize superconductivity in FeSe (e.g., by suppressing a nematic order), they are not sufficient conditions for the occurrence of high-$T_{\rm c}$ superconductivity in FeS. In other words, the pairing interaction inherent in the electron-doped FeSe layer is much stronger than that in the electron-doped FeS layer. Then, the important question is what kind of interaction in FeSe layer triggers superconductivity. To clarify this point, we compared the band dispersions between 1-ML FeS and FeSe at a similar doping level and found a clear difference in the $\gamma$-band width (Fig. 3 F and G). The band width is twice larger in FeS than in FeSe although the both bands originate from the same Fe-3$d$ orbital, but not from the chalcogen $p$ orbital (note that a similar difference in the band width is observed in multilayer films). This result highlights the strongly-correlated nature of FeSe, and implies that the superconductivity in the absence of interfacial electron-phonon coupling may be associated with spin and/or orbital fluctuations enhanced by the strong electron correlation. The strong correlation in FeSe may be also related to the emergence of a Mott-like insulating phase in non-doped 1-ML FeSe (29). In order to get a deeper insight into this issue, the growth of non-doped 1-ML FeS and its ARPES study are important in future.

The absence of high-$T_{\rm c}$ superconductivity in 1-ML FeS which exhibits a clear signature of replica band also puts a strong constraint on microscopic theories to explain how the interfacial electron-phonon coupling contributes to high-$T_{\rm c}$ superconductivity. While the experimental evidence for the importance of interfacial electron-phonon coupling for the superconducting pairing is accumulated thus far, there are controversies among the theoretical proposals regarding the magnitude of $T_{\rm c}$ enhancement; some theories have proposed that the electron-phonon coupling alone explains high-$T_{\rm c}$ superconductivity (16,18), while others suggested that a cooperation of interfacial electron-phonon coupling with other primary pairing mechanisms (likely spin/orbital fluctuations) is essential to achieve high $T_{\rm c}$ (13-15). The absence of high $T_{\rm c}$ in monolayer FeS demonstrates that the interfacial electron-phonon coupling is not strong enough to induce high-$T_{\rm c}$ superconductivity by itself, supporting the latter scenario. The electron-phonon coupling seems to enhance $T_{\rm c}$ only when it cooperates with the pairing interaction inherent to the superconducting layer. This result provides a useful strategy for exploring new atomic-layer high-$T_{\rm c}$ systems; we should at first choose starting materials that inherently have a potential for high-$T_{\rm c}$ superconductivity and then increase $T_{\rm c}$ by using the cooperation with interfacial effects.

In summary, we have succeeded in synthesizing a high-quality 1-ML film of metastable tetragonal FeS by the combined method of the topotactic reaction and MBE. The comparative ARPES measurements between FeS and FeSe have revealed marked similarities (electron doping and electron-phonon coupling across the interface) and differences (presence/absence of high-$T_{\rm c}$ superconductivity and electron correlation strength). The present results clearly show that the excess electron doping and the electron-phonon coupling across the interface may be necessary conditions, but are not sufficient conditions for high-$T_{\rm c}$ superconductivity. The remarkable difference in the $\gamma$ band at $E_{\rm F}$ between FeS and FeSe suggests an important role of strong correlation for high $T_{\rm c}$. The present study has also established the high usefulness of topotactic reaction to synthesize metastable two-dimensional materials. This method would be applied to a wide range of novel two-dimensional materials such as transition-metal dichalcogenides and topological materials.

\section*{Materials and Methods}
A FeTe thin film was grown on a Nb(0.05wt$\%$)-doped SrTiO$_3$ substrate (Shinkosha) with the MBE method in a vacuum better than 2 $\times$ 10$^{-10}$ Torr. The substrate was first degassed at 600$^{\circ}$C for 2 h, and then annealed at 900$^{\circ}$C for 30 min. A FeTe film was grown by co-evaporating Fe and Te in a Te rich condition while keeping the substrate temperature at 270$^{\circ}$C. The film thickness was controlled by varying the deposition time with keeping the constant deposition rate (approximately 0.01 ML/sec). After the evaporation, the film was post-annealed at 300$^{\circ}$C for 2 h in a vacuum. Next, the film was heated to 270$^{\circ}$C and exposed to a sulfur (S) molecular beam (the partial pressure of $\sim$10$^{-8}$ Torr) for 5 min to obtain a FeS film. Then, the FeS film was post-annealed at 530$^{\circ}$C for 1h in a vacuum better than 2 $\times$ 10$^{-10}$ Torr. After the growth, the film was immediately transferred to the ARPES-measurement chamber which is directly connected to the MBE chamber. Reflection high-energy electron diffraction (RHEED) and low-energy electron diffraction (LEED) measurements were employed to characterize the surface cleanness and structure of the sample. Energy-dispersive x-ray spectroscopy (EDX) and x-ray photoemission spectroscopy (XPS) measurements were performed to confirm the substitution of Te with S atoms. The spatial resolution in the present EDX setup ($\sim$50-nm sample thickness and 200-keV incident beam energy) is estimated to be about 2 nm by referring to the textbook (30). Deposition of cesium (Cs) atoms was carried out at room temperature with a Cs dispenser (SAES Getters).

ARPES measurements were performed in an ultrahigh vacuum better than 5 $\times$ 10$^{-11}$ Torr using a Scienta-Omicron SES2002 spectrometer with a high-flux He discharge lamp ($h\nu$ = 21.218 eV) at Tohoku University. The energy and angular resolutions were set to be 2-30 meV and 0.2$^{\circ}$, respectively. The Fermi level of sample was referenced to that of a gold film evaporated onto the sample holder.

\begin{acknowledgments}
We thank K. Fujiwara for discussion. This work was supported by Japan Society for the Promotion of Science KAKENHI Grants JP17H04847, JP17H01139, JP18H01160, JP18K18986, and JP18H01821; The Ministry of Education, Culture, Sports, Science and Technology of Japan (In- novative Area ``Topological Materials Science" Grant JP15H05853); the Japan Science and Technology Agency Precursory Research for Embryonic Science and Technology Grant JPMJPR18L7; Core Research for Evolutional Science and Technology Grant JPMJCR18T1; and High Energy Accelerator Research Organization (KEK) Photon Factory Proposal 2018S2-001.
\end{acknowledgments}

\bibliographystyle{prsty}

\end{document}